\documentclass[conference,compsoc]{IEEEtran}

\usepackage{cite}
\usepackage[pdftex]{graphicx}
\usepackage{amsmath}
\usepackage{flushend}
\usepackage{url}
\usepackage{svg}
\usepackage{todonotes}

\begin{document}
\title{
    Testing the Accuracy of Surface Code Decoders
}

% author names and affiliations
% use a multiple column layout for up to three different
% affiliations
\author{\IEEEauthorblockN{Arshpreet Singh Maan}
\IEEEauthorblockA{Aalto University, Espoo, Finland\\
arshpreet.maan@aalto.fi}
\and
\IEEEauthorblockN{Alexandru Paler}
\IEEEauthorblockA{Aalto University, Espoo, Finland\\
alexandru.paler@aalto.fi}
}

\maketitle

\begin{abstract}
Large-scale, fault-tolerant quantum computations will be enabled by quantum error-correcting codes (QECC). This work presents the first systematic technique to test the accuracy and effectiveness of different QECC decoding schemes by comparing a look-up table decoder to solutions generated using algorithmic decoders. Specifically, we examine the results of minimum-weight-perfect-matching and belief-propagation decoders against exhaustive look-up tables for surface codes up to distance seven and categorise where errors are accurately corrected in both decoding schemes.  While our results are preliminary, we show that significant quantitative results can be generated, comparing how actual error channels are successfully or unsuccessfully decoded.  We show that different decoding schemes perform very differently under the same QECC scheme and error model, and detail how decoders can be tested and classified with respect to errors that are successfully decodable.  This work paves the way to the data driven tuning of decoder ensembles and will enable tailored design of hybrid decoding schemes that allow for real-time decoding, while maintaining the high theoretical thresholds allowed by specific quantum error correction codes.
\end{abstract}

\IEEEpeerreviewmaketitle

\section{Introduction}

Fault-tolerant quantum computations require hardware that fails with a probability below a value called threshold failure rate, and software that can detect and correct, as fast and as accurate as possible, the hardware failures. To protect the qubit states, error correction is a necessity and various quantum error correcting codes (QECC) have been proposed so far. A QECC uses data qubits and syndrome qubits. The measurement of the syndrome qubits is used to detect errors: the state of the syndrome qubit is flipped if some error patterns happen on the associated data qubits. Assuming that measurements are perfect, a flipped syndrome qubit's state is indicative of potential errors on the data qubits. A decoder is an algorithm that takes as input a string of syndrome bits, and outputs the most probable error on the data qubits. The threshold of a QECC is the hardware failure rate at which the benefits of the code start to be visible: exponentially low failure rates of the encoded quantum state are achieved with polynomial overheads (qubits and time).

Recent demonstrations of error-corrected quantum computations, such as~\cite{google2023suppressing,Krinner_2022, surface_code_exp} were enabled by the improvements (e.g. gate fidelity, coherence times and qubit control) and better understanding of the hardware failures, as well as the development of software methods for supporting the error correction (e.g. a recent review of QECC decoders is~\cite{iOlius_2022}).

There seem to be two options to achieve error-corrected quantum computations at scale: 1) significantly lowering the hardware failure rates; 2) improving the QECC software to tolerate high hardware error rates. While the first is a very desirable goal, it is an extremely complex task. The latter is challenging too, but it is considered to be more realistic. High-threshold, error-corrected computations are possible with the surface QECC~\cite{fowler2012surface} and the appropriate decoder.

\begin{figure}[!t]
\centering
    \includegraphics[width=\columnwidth]{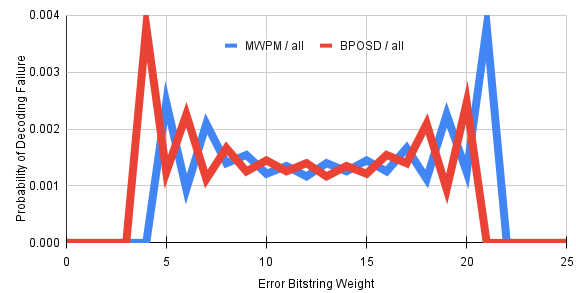}
    \caption{The failure rate distribution for surface code distance five. The horizontal axis represents the weight of the bit string representing the error. The failure rates are computed for the errors that were wrongly decoded by only one of the decoders. For example, the blue MWPM line represents the distribution of the errors which were wrongly decoded by MWPM, but were decoded correctly by BPOSD. For low weight errors BPOSD fails more often compared to MWPM. High weight errors are not realistic, because they are significantly above the threshold of the surface code. We conclude that the better threshold of MWPM compared to BPOSD is explained by the very low weight errors that are difficult for BPOSD. The decoders perform approximately equally well once the error weight increases.}
    \label{fig:comparison_ratio}
\end{figure}

We are motivated to finding out \emph{why}, for the case of surface codes, a particular decoder performs better than the other. Which errors can be corrected by one or the other decoder? Are there patterns in the errors that one of the decoder can handle? Our work is motivated by the goal of achieving fast high-threshold decoding. In parallel to our work, the authors of~\cite{delfosse2023choose} discuss a method for choosing the best decoder based on the speed vs. accuracy trade-off.

\begin{figure*}
    \centering
    \includegraphics[width=0.20\textwidth]{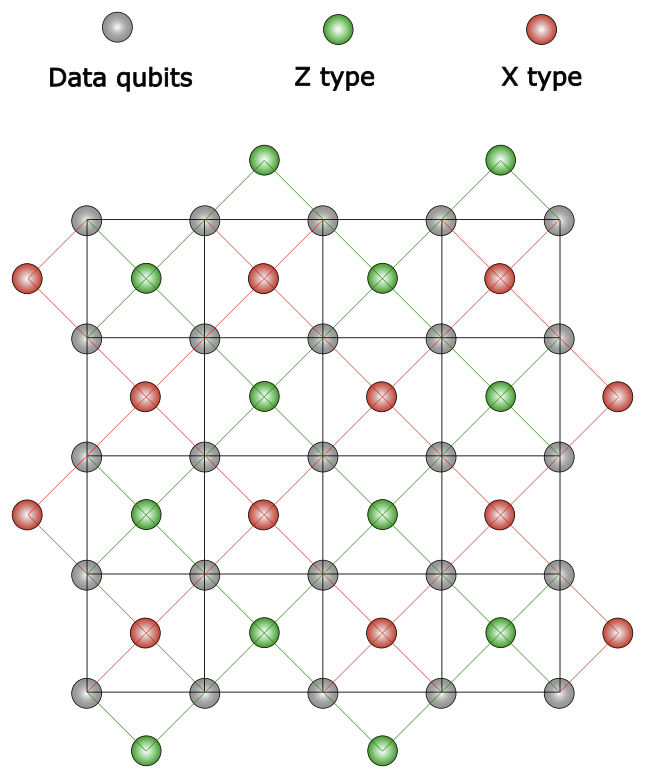}
    \includegraphics[width=0.20\textwidth]{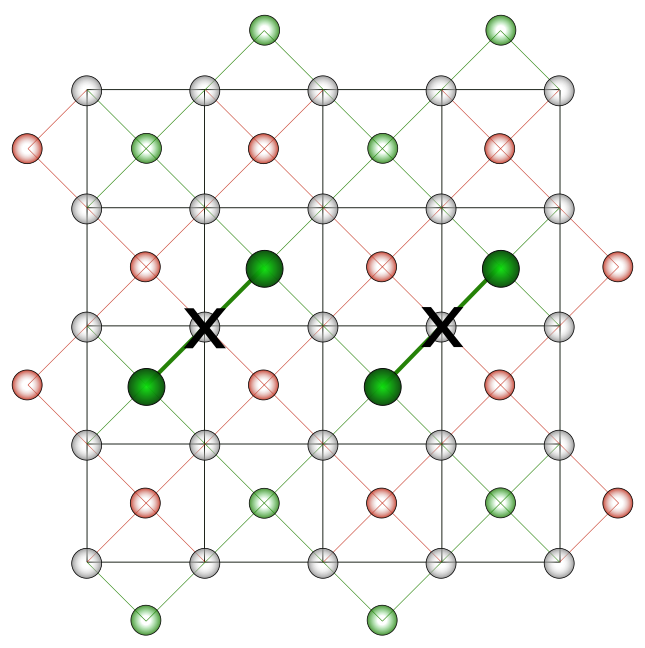}
    \includegraphics[width=0.20\textwidth]{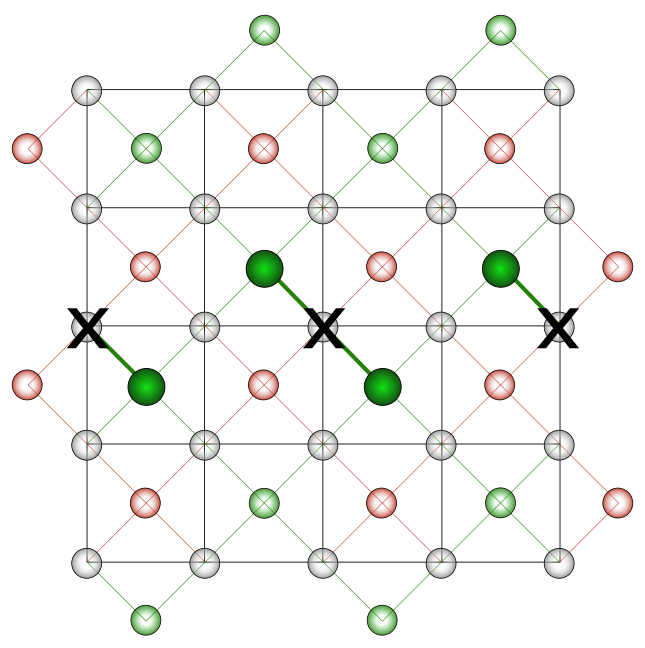}
    
    \caption{An example of a difficult error pattern is shown on a surface code illustrated in a). The difficulty arises due to the code degeneracy: the same syndrome bits (red and green vertices) can be generated by different errors on data qubits (black vertices). There are two types of difficult syndromes: one which a decoder should theoretically be able to decode but fails to do so (in (b)) and one which decoder should theoretically fail, but it somehow decodes (in (c)); b) Two data qubits on the third row (black vertices) encounter an X error, and four syndrome qubits (green) are flipped; c) The same four syndrome qubits could have been flipped if the other three data qubits on the third row would have had X errors. The error in c) is more unlikely than the one in b) -- and most of the decoders, including MWPM, will do their best to predict the error pattern from b). Assumming that the correct error pattern is the one in b), a decoder that predicts the pattern from c) will silently introduce a logical error -- the original two errors plus the wrong three corrections form a chain that is connecting the two boundaries of the surface code patch, and this is effectively performing a logical operator on the encoded state. This unwanted decoding behaviour is lowering the accuracy of the decoder.}
    \label{fig:example}
\end{figure*}

\subsection{Background}

Any QECC needs to protect only against bit flip errors (modelled by the Pauli X operator) and phase flip errors (modelled by the Pauli Z operator). Any other potential error is a linear combination of these ones. A surface code is defined on a two dimensional square lattice~\cite{fowler2012surface}.

For the case of the surface code, two different syndrome extraction circuits are repeatedly applied to the data qubits in order to extract information about X and Z errors from neighbourhoods of four data qubits. Each syndrome extraction round is called a \emph{surface code cycle}.

In general, the weight of a bit string is the number of bit $1$ in the string. In particular, for a distance $d$ surface code, there are $(d^2 - 1)$ syndrome bits extracted in each cycle, and the weight can be any value between zero and $(d^2 - 1)$. At the same time, there are $d^2$ error locations and, in the following, when we refer to error bit strings we mean a sequence of $d^2$ bits which indicate if an error happened on a particular data qubit. This is distinct from the \emph{weight of error}, which is the number of errors from the \emph{error bit string} that can corrupt the encoded state.

QECC decoding is hindered by the degeneracy of the QECC~\cite{degeneracy, poulin2008iterative}: many different errors generate the same syndrome. Most of the time, the corrections do not need to be applied immediately to the data qubits, and can instead be stored and tracked in a so called Pauli frame, e.g. \cite{paler2014software}. The results stored within the frame are used when reading out the error-corrected data.

When benchmarking decoders, or computing thresholds, it is extremely challenging to model and simulate the quantum noise sources~\cite{lidar_brun_2013,karimi2023qubit} precisely, such that simplified models are usually employed. The \emph{code capacity noise model} assumes Pauli errors on data qubits happen with some probability $p$, while the syndrome qubits and their measurements are assumed perfect. The \emph{phenomenological noise model} considers Pauli errors as well as measurement errors on both data and syndromes qubits with some probability $p$. The \emph{circuit level noise model} model assumes everything is faulty at the circuit level, the gates, the wires as well as the measurements. A single Pauli error at the circuit level can be transformed into multiple Pauli errors due to the presence of CNOT gates in the syndrome extraction circuit. In this work we focus on the code capacity model.

Multiple decoding algorithms~\cite{iolius2023decoding, Delfosse2021almostlineartime, Varsamopoulos_2017} have been proposed for surface codes. Herein, we use the Minimum Weight Perfect Matching decoder (MWPM)~\cite{higgott2021pymatching, higgott2023sparse} and the Belief Propagation - Ordered Statistics Decoder (BPOSD)\cite{panteleev2022asymptotically, Roffe_2020}.

MWPM takes as input an error detection graph representing the syndromes (detection events) and performs perfect matching on the graph such that it minimizes the weight of the matching~\cite{fowler2012surface}. The belief propagation~\cite{pearl1988probabilistic} algorithm works by exchanging messages between syndrome nodes and data nodes on the Tanner graph, which is a graphical representation of the error correcting code. BP tries to find the qubit-wise most likely error \cite{degeneracy,poulin2008iterative} and converges when the predicted syndrome equals the actual syndrome. BP is not optimal in the degenerate QECC case and can be improved by post-processing such as Ordered Statistics Decoding (OSD)~\cite{panteleev2022asymptotically}.

Decoders have to be fast and accurate at the same time, in general. The worst case complexity of MWPM in the number of detection graph vertices $N$ is $O(N^3\log{N})$, and lower complexity variations of MWPM have been proposed, e.g.~\cite{fowler2014minimum, higgott2023sparse}. BPOSD has a worst case complexity of $O(N^3)$ with most of the complexity dominated by the OSD post-processing~\cite{panteleev2022asymptotically}.

The general rule is that the deadline for decoding each round of input syndromes is, on average, the duration of a code cycle, or otherwise the decoding backlog will grow exponentially~\cite{terhal2015quantum}. When numerically estimating threshold-error rates for various noise models, the observation is that speed is gained at the cost of decoding accuracy. For the surface code, under the code capacity noise model, MWPM offers a threshold of $14\%$~\cite{iolius2023decoding}, and BPOSD has a threshold of $13.9\%$~\cite{iolius2023decoding}.

\subsection{Motivation}

The goal of this work is to develop a methodology for understanding why and in which situations a particular decoder performs better than another one. To the best of our knowledge, our approach has not been pursued by now, although look-up table~\cite{das2021lilliput} decoders have already been implemented. A look-up table decoder is systematically mapping errors to syndrome information.

A QECC of distance $d$ should deterministically detect and correct all errors of weight up to $(d-1)/2$. The \emph{accuracy} of a decoder is its capability to do exactly that. However, due to the degeneracy of the QECCs and various decoder configuration parameters, accuracy is not guaranteed. There are two effects regarding decoder accuracy: a) coincidentally high weight errors that happen to be correctable; b) low weight errors which should be correctable but the decoder fails. Both effects impact the threshold, but only the latter impacts the accuracy of the decoder.

A \emph{difficult error} is one that cannot be corrected based on the syndrome that it is generating: the decoder corrects the wrong error. We motivate our work by an example: Fig.\ref{fig:example} presents how a difficult error can arise when decoding surface codes. Difficult errors are very common due to the degeneracy of the surface code.

We are also interested in testing the computational limits of building lookup table (LUT) decoders that stores \emph{difficult syndromes}. Such syndromes are corrected by a slow decoder, but not by a fast decoder. Due to code degeneracy, a straightforward syndrome that \emph{often} leads to a correction, and a difficult syndrome is one which is \emph{seldom} leading to a correction. We envision hybrid decoders, where the LUT is combined with a fast decoder that will handle the straightforward syndromes.

Preparing LUT decoders is computationally a very expensive task. However, we will show that up to distance nine ($9 \times 9 = 81$ data qubits), this can be achieved by exhaustive enumeration on a supercomputer the size of LUMI\cite{zwinger2023lumi}. The 81 qubits of a distance nine surface code can be used to accommodate nine distance three patches ($3 \times 3$ qubits). The patches can be interacted by lattice surgery~\cite{horsman2012surface}. Therefore, LUT decoding could be the first path to achieving fast and efficient decoding close to the threshold error rates of the surface code.

%\subsection{Contribution}

\section{Methods}

\begin{figure}[!t]
    \centering
    \includegraphics[width=\columnwidth]{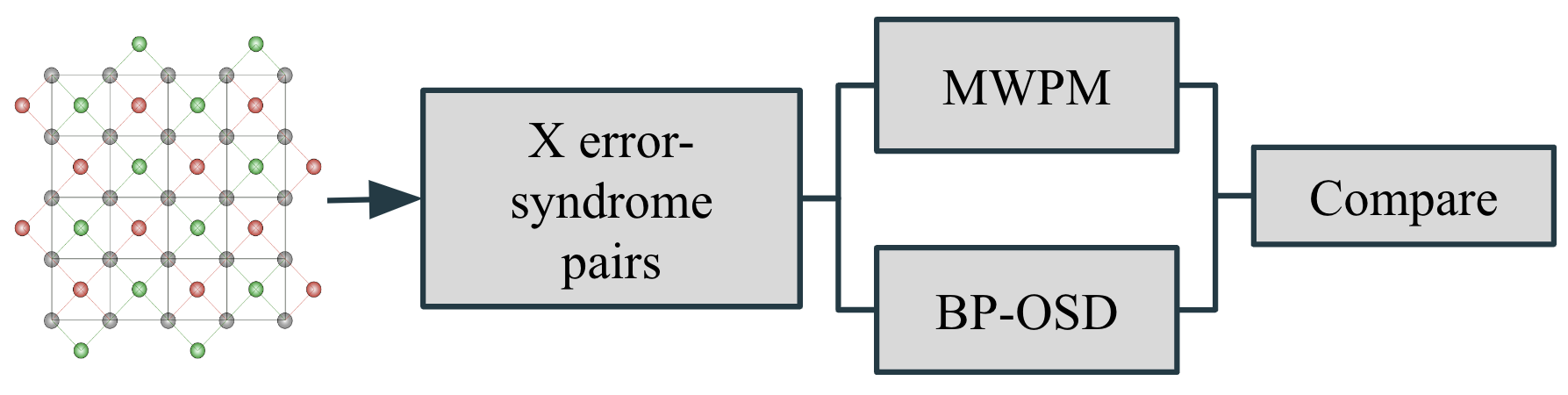}
    \caption{Workflow for comparing decoders. 1) Error bit strings and corresponding syndrome bit strings are generated starting from a distance $d$ surface code patch; 2) The same syndrome is sent to each decoder (MWPM and BPOSD); 3) Each decoder outputs the most probable correction and the corrections are applied to the errors we generated in the first step; 4) The decoders are compared by computing if the corrections resulted in a logical error or not.}
    \label{fig:schematic}
\end{figure}

Our method for determining difficult errors is based on exhaustive simulation. In the following, we assume the code capacity error model and focus solely on single qubit bit flip errors. Under code capacity noise, for a distance $d$ rotated surface code, there will be $2^{d^2}$ possible errors and $2^{{(d^2-1)}}$ possible syndromes. However, because we consider only bit flips, half of the syndrome qubits, i.e. the phase-flip syndrome checks will not be enabled, and there will be $2^{{(d^2-1)}/2}$ distinct syndromes.

\subsection{Exhaustive Search}

We enumerate all bit strings of the numbers between zero and $2^{d^2}$, and compute the syndrome bit strings pattern using the surface code parity check matrix. We generate all error-syndrome pairs for distances three and five. For distance seven we compute only a subset of these pairs.

Our method, illustrated in Fig.~\ref{fig:schematic}, follows the traditional procedure for the numerical calculation of code thresholds. However, we use two decoders in order to find out which errors are \emph{difficult} for one or the other decoder, or both. We generate all error-syndrome pairs and send these pairs to the MWPM and BPOSD decoders, to predict the most probable correction. The predicted corrections are applied on to the actual error resulting into the \emph{residual error}. A logical error occurs if the residual error changes the logical state of the surface code, or if the weight of the corresponding syndrome bit string is higher than zero.

We store the weight of the error for which the respective decoder failed. Consequently, we are storing three different failure weights: 1) for which MWPM failed but BPOSD succeeded, 2) for which BPOSD failed but MWPM succeeded, and 3) the weight of the error for which both decoders failed. %We further analyse these weights for any patterns in the failures of a respective decoder.

\begin{figure*}
\centering
    \includegraphics[width=0.4\textwidth]{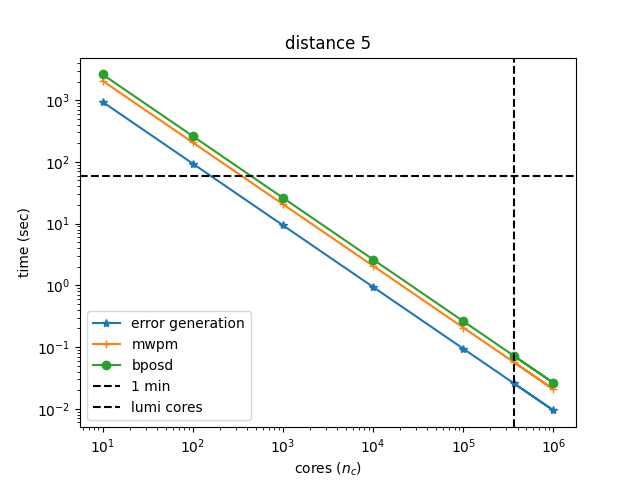}
    \includegraphics[width=0.4\textwidth]{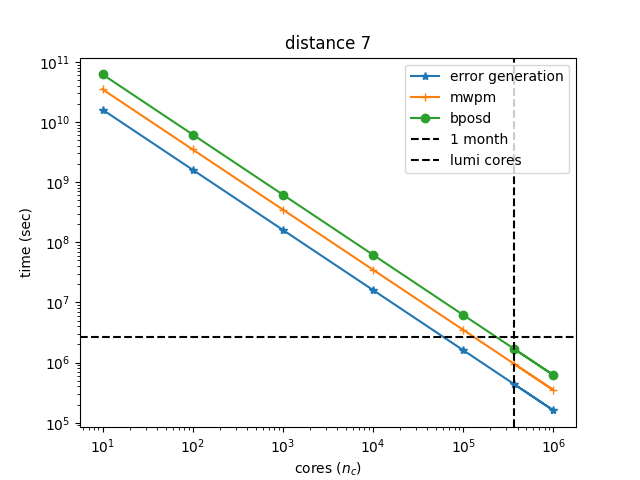}
\caption{The estimated time taken by MWPM and BPOSD to decode exhaustively all possible errors for: a) distance 5; b) distance 7. The vertical line is the number of cores in the LUMI supercomputer.}
\label{fig:results_time}
\end{figure*}

Our method has exponential complexity and will not scale for large distance codes. However, we can predict the maximum distance for which it is feasible (Fig.~\ref{fig:results_time}). We know that with the help of a supercomputer it is possible to pre-compute all possible error-corrections in a LUT decoder, in order to enable distance seven and nine real-time decoding experiments with real quantum hardware.

\subsection{Feasibility Study}

To predict the feasibility of exhaustively analysing the largest possible decode-able distance, we assume the same code capacity noise model. We sample random bit strings to average the decoding time for low weight (weight $ < d/2$) and high weight (weight $\geq d/2$) errors. We generate 1 million random bit strings, between zero and $2^{d^2}$, representing the error and generate the corresponding syndrome. For the 1 million errors, we measure on a single CPU core three separate total times $t_i$: 1) the time taken to generate; 2) the time needed to decode by MWPM, and 3) the time needed to decode by BPOSD. We average the time per syndrome, and estimate afterwards the time it would take to generate and decode, using a single core, exhaustively distance $d \in \{5, 7, 9\}$ on a single core by $T_i = (t_i * 2^{d^2})/10^6$.

Our testing method is trivially parallel, and we are interested in finding out if multi-core computations could increase its feasibility. We \emph{assume} that we split our dataset on $n_c$ number of cores\footnote{Finland's super computing cluster $LUMI$ has $362,496$ cores~\cite{zwinger2023lumi}.} and predict a worst case execution time $T_m = T_i/n_c$. Results are presented in Fig.~\ref{fig:results_time}.

The code used for our analysis was not fully optimized for performance (pure Python code that does not use vectorized CPU instructions, efficient register allocation strategies and caching). However, using a highly optimized code, one can decode up to 1 million errors per core second~\cite{higgott2023sparse}. We expect orders of magnitude reduction in decoding time on high performance cores of supercomputers, such that distance 9 codes can also be tested in practice. 

\section{Results}

\begin{figure*}
    \centering
    \includegraphics[width=0.39\textwidth]{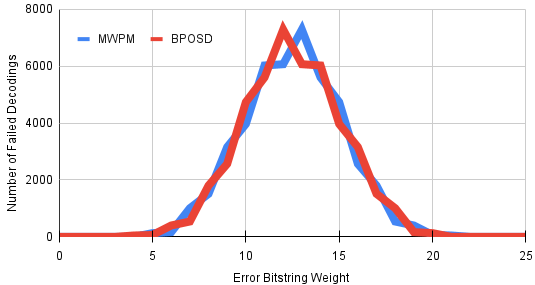}
    \includegraphics[width=0.39\textwidth]{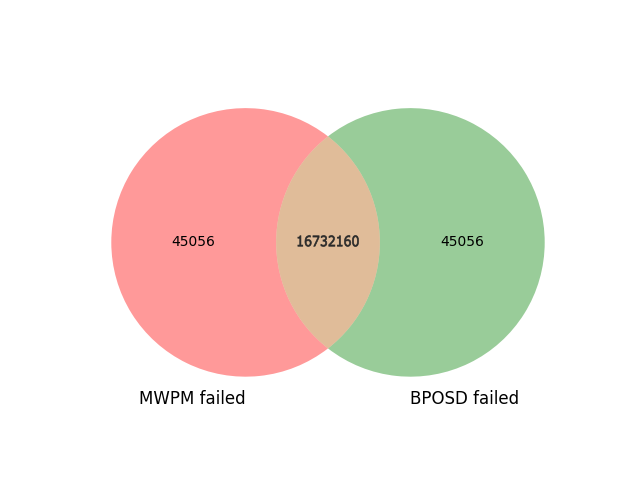}

    \caption{left)(a) The plot shows, for distance five the distribution of the weight of the error for set A - A $\cap$ B, where A = decoder A failed, B = decoder B failed, A $\cap$ B = both decoders failed;  
    right)(b) The Venn diagram (not to scale) shows for distance five, out of all possible error-syndrome pairs, the number of the errors where MWPM failed but BPOSD succeeded (red), BPOSD failed but MWPM succeeded (green) and where both failed i.e. MWPM failed $\cap$ BPOSD failed.}
    \label{fig:venn_hist}
\end{figure*}

The results of exhaustively simulating all the errors and sending the corresponding syndromes to the MWPM and BPOSD decoders are shown in Fig.~\ref{fig:comparison_ratio} and ~\ref{fig:venn_hist}. We used \emph{panqec} \cite{huang2023tailoring} for preparing our experiments, and ran the experiments on an M2 Macbook Pro with 16GB of memory. Table~\ref{tab:config} contains the parameters of the experiments and the time it would take to run the entire experiment on LUMI. The probability of error is theoretically not necessary when enumerating through all the error bit strings, but the panqec decoders required it for initialisation.

\begin{table}[!h]
    \centering
    \caption{Decoding Experiment}
    \label{tab:config}
    \begin{tabular}{c|c|c|c|c}
    Dist& Nr. errors    & Decoder   & Prob. & Time LUMI (s)\\
    \hline
    5   &   33554432    & MWPM      & 0.1   &  5.70e-02\\
        &               & BPOSD     & 0.1   &  7.23e-02\\
    \hline
    7   &   5.6295e+14  & MWPM      & 0.1   &  9.62e+05\\
        &               & BPOSD     & 0.1   &  1.70e+06\\
   \hline
    9   &   2.4179e+24   & MWPM      & 0.1   &  4.19e+15\\
        &               & BPOSD     & 0.1   &  9.98e+15\\
    
    \end{tabular}
\end{table}

The Venn diagram in Fig.~\ref{fig:venn_hist}(b) shows for distance five, out of all possible error-syndrome pairs, the number of the errors where MWPM failed but BPOSD succeeded (red), BPOSD failed but MWPM succeeded (green) and where both failed i.e. MWPM failed $\cap$ BPOSD failed.  The diagram shows that both decoders, in total, fail for the same number of errors. However, when we look at the distribution of the weight of the error, it presents a different picture as shown in Fig.~\ref{fig:venn_hist}(a). The plot shows, for distance five the distribution of the weight of the error for set A - A $\cap$ B, where A = decoder A failed, B = decoder B failed, A $\cap$ B = both decoders failed. Fig.~\ref{fig:comparison_ratio} shows the distribution of ratio of failure of set A - A $\cap$ B  with respect to all possible errors for a given weight. We observe that the MWPM performs better than BPOSD for low weight error bit strings, while BPOSD performs better than MWPM for high weight error bit strings.

\section{Conclusion}

We presented a method for testing the accuracy of a QECC decoder: are the errors expected to be corrected by a decoder also corrected? We test by comparing a decoder under test with a look-up table decoder (that has the mapping between error and syndromes hardwired). Our method is computationally expensive, but we show that, if supercomputing capabilities are available, it is feasible to apply it to surface codes up to distance nine. We analyze two distinct decoders (MWPM and BPOSD) and show how the decoding algorithms are sensitive to the errors that generate the syndromes. Future work will focus on a more complex analysis of the error channels in order to enable the data driven tuning of decoders.

\section*{Acknowledgment}
This research was developed in part with funding from the Defense Advanced Research Projects Agency [under the Quantum Benchmarking
(QB) program under award no. HR00112230007 and HR001121S0026 contracts], and was supported by the QuantERA grant EQUIP through the Academy of Finland, decision number 352188. We thank Simon Devitt for the discussions about the method and the significant improvements to the manuscript's clarity, Karoliina Oksanen and Mikko Seesto for the initial code used herein, and Niki Loppi of the NVIDIA AI Technology Center Finland for his help with the implementation.

\bibliographystyle{IEEEtran}
\bibliography{__main}
\end{document}